\documentclass[showpacs,aps,pre,twocolumn,amsmath,amssymb,epsfig]{revtex4-1}

\usepackage{graphicx}
\usepackage{epsfig}
\usepackage{dcolumn}
\usepackage{bm}
\newcommand     {\beq}[1]         { \begin{equation} #1 \end{equation} }

\begin{document}

\title{System size dependent avalanche statistics in the limit of high disorder}

 \author{Vikt\'oria K\'ad\'ar}
  \author{Ferenc Kun}
 \email{Corresponding author: ferenc.kun@science.unideb.hu}
  \affiliation{Department of Theoretical Physics, University of Debrecen,
 P.O. Box 5, H-4010 Debrecen, Hungary}
 \affiliation{Institute for Nuclear Research, Hungarian Academy of Sciences 
  (Atomki), P.O. Box 51, H-4001 Debrecen, Hungary}

\begin{abstract}
We investigate the effect of the amount of disorder on the statistics of breaking 
bursts during the quasi-static fracture of heterogeneous materials.
We consider a fiber bundle model where the strength of single fibers is sampled 
from a power law distribution over a finite range, so that the amount 
of materials' disorder can be controlled by varying the power law exponent and the upper cutoff 
of fibers' strength. 
Analytical calculations and computer simulations, performed in the limit of equal load sharing,
revealed that depending on the disorder parameters the mechanical response of the bundle 
is either perfectly brittle where the first fiber breaking triggers a catastrophic 
avalanche, or it is quasi-brittle where macroscopic failure is preceded by a sequence of bursts.
In the quasi-brittle phase, the statistics of avalanche sizes is found to show
a high degree of complexity. In particular, we demonstrate that the functional form 
of the size distribution of bursts depends on the system size: for large upper cutoffs of fibers' strength, 
in small systems the sequence of bursts has a high degree of stationarity characterized by 
a power law size distribution with a universal exponent. 
However, for sufficiently large bundles the breaking process accelerates towards 
the critical point of failure which gives rise to a crossover between two 
power laws. The transition between the two regimes occurs at a characteristic system size
which depends on the disorder parameters.
\end{abstract}

\date{\today}

\maketitle

\section{Introduction}
The disorder of materials plays a crucial role in fracture phenomena when subject to
mechanical loads. Experiments and theoretical calculations revealed that 
under constant or slowly 
varying external loads the fracture of heterogeneous materials proceeds in bursts of 
local breakings \cite{petri_experimental_1994,maes_criticality_1998,deschanel_experimental_2009,
vives_0953-8984-25-29-292202,vives_coal_burst_2019,vives_burst_bone_2016}. 
Such crackling events can be recorded in the form of acoustic signals 
providing insight into the microscopic dynamics of the fracture process 
\cite{diodati_acoustic_1991,Lockner1993883,meinders_scaling_2008,niccolini_acoustic_2011}.
Cracking bursts can be considered as precursors of the ultimate 
failure of the system, so that they can be exploited to forecast the impending catastrophic event
\cite{vasseur_scirep_2015,salje_main_minecollapse_2017,alava_lifetime_pre2016,
pradhan_crossover_2005-1,davidsen_scaling_2007,sornette_predictability_2002,niccolini_acoustic_2011,
voight_ffm_nature_1988,main_limits_2013}.

The intensity of the crackling activity has been found to depend 
on the degree of materials' disorder
\cite{rosti_crackling_2009,vasseur_scirep_2015,vives_coal_structure_2019}: in the limiting 
case of zero disorder, the ultimate failure occurs in an abrupt way with 
hardly any precursory activity \cite{zapperi_first-order_1997,picallo_brittle_2010}.  
However, the presence of disorder gives rise to a gradual cracking process where macroscopic 
failure occurs as a result of the intermittent steps of damage accumulation 
\cite{guarino_experimental_1998,ramos_prl_2013,santucci_sub-critical_2004}.
Recently, experiments have been performed on the compressive failure of porous glass samples
where the degree of heterogeneity could be well controlled during the sample preparation 
\cite{vasseur_scirep_2015}. 
These experiments have shown that increasing disorder gives rise 
to a more intensive bursting activity 
with a higher number of cracking events whose size spans a broader range. 
As a consequence, the precision 
of failure forecast methods was found to improve with increasing disorder 
\cite{vasseur_scirep_2015}.

Motivated by these recent findings, here our goal is to investigate the statistics 
of crackling noise in the limiting case of extremely high disorder. 
The fiber bundle model (FBM) provides an adequate framework 
\cite{de_arcangelis_scaling_1989,andersen_tricritical_1997,hansen2015fiber,
kloster_burst_1997,kun_extensions_2006,hidalgo_universality_2008,
hidalgo_avalanche_2009} to study the statistics of breaking avalanches allowing for 
a simple way to control the degree of disorder \cite{biswas_lls_2017,ray_epl_2015,
hidalgo_universality_2008,PhysRevE.87.042816,udi_epl_2011,materials_fiber_2017}. 
In FBMs the sample is discretized in terms of parallel fibers
where controlling the mechanical response, strength and interaction of fibers various
types of materials' behaviours can be captured. Disorder can be represented 
by the random strength of fibers while their Young modulus is kept constant.
In our study, high disorder is realized by a power law distribution 
of fibers' strength over a finite range 
where the amount of disorder can be controlled by the exponent and by the upper cutoff 
of the strength values.

Assuming equal load sharing after fiber breakings, 
we demonstrate that the fat tailed microscale disorder has a substantial effect 
on the statistics of breaking bursts of fibers. 
In particular, we show that the functional form of the burst size distribution 
depends on the size of the bundle: when the upper cutoff of fibers' strength is infinite 
the burst size distribution is a power law with a universal exponent. However, 
in the case of finite upper 
cutoff strength, for small system sizes the size distribution is identical with 
the one of the infinite cutoff strength. Deviations start at a characteristic system size
beyond which a crossover occurs to another functional form. We give an explanation
of the system size dependent avalanche statistics in terms of the extreme order statistics 
of breaking thresholds.

\section{Fiber bundle model with fat-tailed disorder}
We consider a bundle of $N$ parallel fibers, which are 
assumed to have a perfectly brittle behavior with a Young's modulus $E$ 
and breaking threshold $\sigma_{th}$. The Young's modulus is assumed to be 
constant $E=1$ so that materials' disorder is captured 
by the randomness of the breaking threshold $\sigma_{th}$.
The strength of individual fibers $\sigma_{th}^i$, $i=1,\ldots , N$ 
is sampled from a probability density $p(\sigma_{th})$. The amount of disorder 
in the system can be controlled by varying the range $\sigma_{th}^{min}\leq \sigma_{th} \leq 
\sigma_{th}^{max}$ of strength values and by the functional form of $p(\sigma_{th})$. FBMs 
with moderate amount of disorder have been widely studied in the literature considering 
uniform, Weibull, and Gaussian distributions making the avalanche statistics of this 
universality class well understood \cite{hidalgo_avalanche_2009,hansen2015fiber}.

To realize the limiting case of extremely high disorder, 
a fat tailed disorder distribution is considered, i.e.\ we implement
a power law distribution of breaking thresholds over the range 
$\sigma_{th}^{min}\leq\sigma_{th}\leq\sigma_{th}^{max}$
with the probability density
\begin{eqnarray}
\ p(\sigma_{th}) = \left\{  
\begin{array}{ccc}
0, &  \sigma_{th}< \sigma_{th}^{min},\\ [2mm]
\ A \sigma_{th}^{-(1+\mu)}, &  \sigma_{th}^{min}\leq\sigma_{th}\leq\sigma_{th}^{max}, \\ [2mm]
0, &  \sigma_{th}^{max}<\sigma_{th}. \\
\end{array}
\right.            
\label{surusegfv}
\end{eqnarray}
In our calculations, the lower bound of thresholds $\sigma_{th}^{min}$ is fixed 
to $\sigma_{th}^{min}=1$, while the amount of disorder is controlled by varying 
the power law exponent $\mu$ and the upper bound $\sigma_{th}^{max}$ of thresholds.
The value of $\sigma_{th}^{max}$ covers the range $\sigma_{th}^{min}\leq\sigma_{th}^{max}\leq+\infty$, 
while the disorder exponent is varied in the interval $0\leq\mu< 1$.
For this choice of $\mu$, in the limiting case of an infinite upper bound $\sigma_{th}^{max}\to \infty$
the thresholds do not have a finite average, hence, varying the two parameters $\mu$ and 
$\sigma_{th}^{max}$ the amount of disorder can be tunned in the bundle 
between the extrems of zero and infinity.
Of course, at finite cutoffs $\sigma_{th}^{max}$, the average fiber strength 
$\left<\sigma_{th}\right>$ is always finite, however, 
the specific values of $\sigma_{th}^{max}$ and $\mu$ have 
a very strong effect on the behavior of the system both on the macro- and micro-scales.
The cumulative distribution $P(\sigma_{th})$ can be obtained from the normalized density as
\begin{equation}
\label{eq:distrib}
P(\sigma_{th})= \left\{  
\begin{array}{cc}
0 & \sigma_{th}<\sigma_{th}^{min},\\[1mm]
{\displaystyle \frac{\sigma_{th}^{-\mu} - (\sigma_{th}^{min})^{-\mu}}{(\sigma_{th}^{max})^{-\mu} - 
(\sigma_{th}^{min})^{-\mu}}}, &  \sigma_{th}^{min}\leq \sigma_{th}\leq \sigma_{th}^{max},\\ [1mm]
1 & \sigma_{th}^{max}<\sigma_{th}.\\
\end{array}
\right.
\end{equation}
\begin{figure}
\begin{center}
\epsfig{file=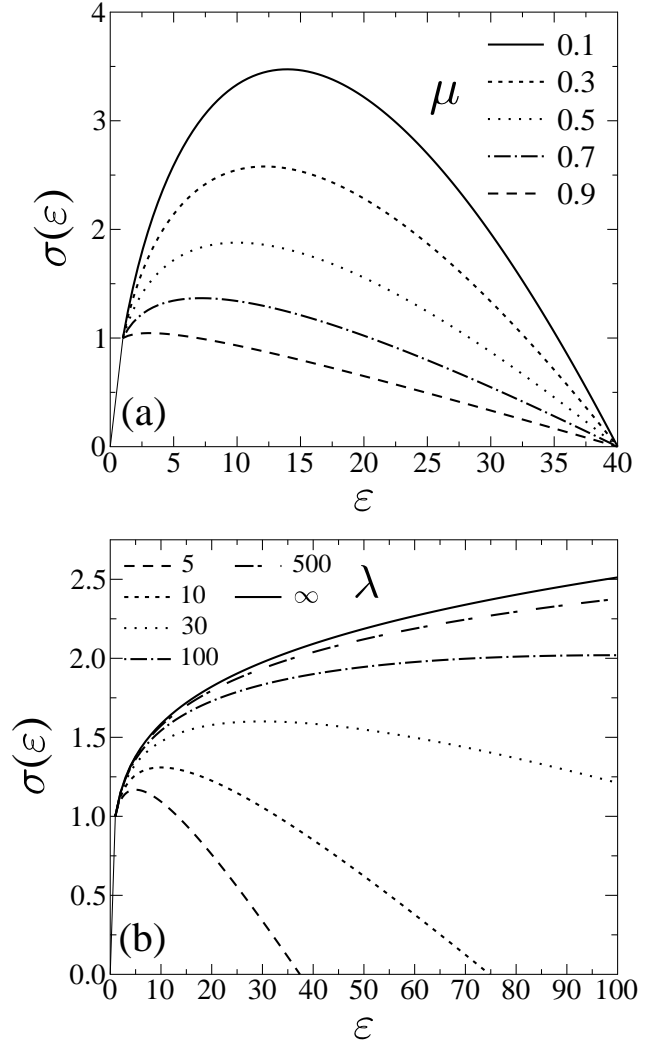,bbllx=30,bblly=40,bburx=380,bbury=650,
width=8.5cm}
\caption{\small Stress-strain curves $\sigma(\varepsilon)$ of the 
bundle $(a)$ for a fixed value of the upper cutoff $\varepsilon_{max}=40$ 
varying the exponent $\mu$, and $(b)$ for a fixed $\mu=0.7$ exponent varying 
the upper cutoff $\varepsilon_{max}$ by means of the multiplication factor
$\lambda$, where $\varepsilon_{max} = \lambda\varepsilon_{max}^c$.
Approaching the phase boundary, in both cases the system becomes 
more and more brittle, i.e.\ the maximum of $\sigma(\varepsilon)$ is preceded by a 
smaller and smaller amount of fiber breakings.
For comparison, the curve corresponding to the case of an infinite upper cutoff 
$\varepsilon_{max}\to \infty$ is also presented.
}
\label{epsilon_sigma_k} 
\end{center}
\end{figure}

After fiber failure, we assume that the excess load of broken fibers is equally redistributed 
over the remaining intact ones. Hence, the constitutive equation
$\sigma(\varepsilon)$ of the bundle can be obtained 
from the general form $\sigma(\varepsilon)=E\varepsilon[1-P(E\varepsilon)]$
by substituting the distribution function $P(x)$ from Eq.\ (\ref{eq:distrib})
\begin{equation}
\label{sigma_th_epsilon_analitikus}
\sigma(\varepsilon)= \left\{
\begin{array}{cc}
\varepsilon, & 0\leq \varepsilon \leq \varepsilon_{min},\\
\displaystyle{\frac{\varepsilon \big(\varepsilon^{-\mu}- \varepsilon_{max}^{-\mu}\big)}{\varepsilon_{min}^{-\mu}-\varepsilon_{max}^{
-\mu}}}, & \varepsilon_{min}\leq \varepsilon\leq \varepsilon_{max},\\
0,& \varepsilon_{max}<\varepsilon.\\
\end{array}
\right.
\end{equation}
For brevity, we introduce the notation $\varepsilon_{min}=\sigma_{th}^{min}/E$, 
$\varepsilon_{max}=\sigma_{th}^{max}/E$, with $E=1$, for the lower and upper bounds 
of strength in terms of strain.
The stress-strain relation of the bundle is illustrated in Fig.\ 
\ref{epsilon_sigma_k}. Perfectly linear behaviour is obtained up to 
the lower bound $\varepsilon_{min}$, since no fibers break in this regime.
After fiber breaking sets on, the constitutive curve becomes gradually non-linear 
and develops a maximum whose position $\varepsilon_c$  and value $\sigma_c$ define 
the tensile strength of the bundle. Both the critical strain $\varepsilon_c$ and stress $\sigma_c$ 
depend on the degree of disorder characterized by $\mu$ and $\varepsilon_{max}$
\begin{equation}
\varepsilon_c = \varepsilon_{max}(1-\mu)^{1/\mu},
\label{eq:crit_strain}
\end{equation}
and
\begin{equation}
\sigma_c= \frac{\mu(1-\mu)^{1/\mu-1}
\varepsilon_{max}^{1-\mu}}{\varepsilon_{min}^{-\mu}-\varepsilon_{max}^{-\mu}}.
\label{eq:sigma_c}
\end{equation}
Recently, we have shown that if the threshold distribution Eq.\ (\ref{surusegfv}) of the 
model 
is sufficiently narrow, already the first fiber breaking can trigger the immediate failure 
of the entire bundle \cite{kadar_pre_2017}. It can be observed in Fig.\ \ref{epsilon_sigma_k} 
that this occurs when the position of the maximum of the constitutive curve $\varepsilon_c$ 
coincides with the lower bound $\varepsilon_{min}$ of the fibers' strength. 
It follows that for all exponent values 
$\mu$ there exists a critical upper bound $\varepsilon_c^{max}$ so that in the range
$\varepsilon_{max} < \varepsilon_{max}^c$ the bundle exhibits a perfectly 
brittle behaviour. 
Perfect brittleness means that under stress or strain controlled loading the breaking 
of the weakest fiber gives rise to an immediate abrupt failure of the bundle, or in a 
softening behaviour, respectively.
The critical upper bound can be obtained from Eqs.\ 
(\ref{eq:crit_strain},\ref{eq:sigma_c}) as
\beq{
\varepsilon_{max}^c=\frac{\varepsilon_{min}}{(1-\mu)^{1/\mu}}.
\label{eq:phase_boundary}
}
The results imply that at a given value of the exponent $\mu$ 
in the parameter regime $\varepsilon_{max} > \varepsilon_{max}^c$
a quasi-brittle response is obtained where macroscopic failure is preceded by breaking avalanches.
\begin{figure}
\begin{center}
\epsfig{bbllx=30,bblly=40,bburx=380,bbury=330, file=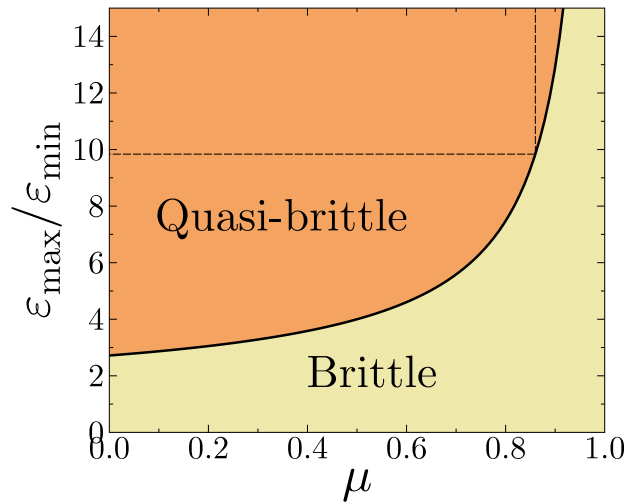, width=8.3cm}
  \caption{Phase diagram of the system. The phase boundary separating the 
  brittle and quasi-brittle macroscopic response is given by Eq.\ (\ref{eq:phase_boundary}).
  Under stress controlled loading, in the brittle phase the bundle suffers immediate abrupt 
  failure at the breaking of the weakest fiber, while in the quasi-brittle phase failure is preceded
  by a sequence of breaking bursts.
  For $\mu\geq 1$ the bundle is always in the brittle phase. The horizontal and vertical dashed lines 
  indicate the parameter sets for which avalanche size distributions were determined by computer simulations.
   \label{fig:crack}}
\end{center}
\end{figure}
The phase boundary separating the brittle and quasi-brittle behaviours of the system 
is given by the relation Eq.\ (\ref{eq:phase_boundary}).
The phase diagram of the system is illustrated in Fig.\ \ref{fig:crack} on the 
$\mu-\varepsilon_{max}$ plane. 
Note that as the exponent $\mu$ approaches 1 from below the value of  
$\varepsilon_{max}^c$ diverges so that the regime $\mu\geq 1$ is always brittle.
When presenting results at a fixed exponent $\mu$, it is instructive to characterize the 
upper cutoff $\varepsilon_{max}$ of fibers strength relative to the corresponding point 
of the phase boundary $\varepsilon_{max}^c(\mu)$. 
Hence, we introduce the parameter $\lambda=\varepsilon_{max}/\varepsilon_{max}^c$, which 
can take any value in the range $\lambda>=1$ (equality holds on the phase boundary between 
the brittle and quasi-brittle phases).

Recently, we have demonstrated that the fat tailed microscale disorder gives rise 
to an anomalous size scaling of the macroscopic strength of the bundle \cite{kadar_pre_2017}. 
For finite upper cutoffs of fibers' strength $\varepsilon_{max}$, 
the average strength of the bundle 
$\left<\varepsilon_c\right>$ was found to increase with the number $N$ of fibers as
\beq{
\left<\varepsilon_c\right> \sim N^{1/\mu}.
\label{eq:strength_size}
}
The usual decreasing behaviour of strength \cite{smith_probability_1980,smith_lower-tail_1983} 
gets restored beyond a characteristic system size $N_c$
which depends on the disorder parameters as
\beq{
N_c\sim \varepsilon_{max}^{\mu}.
\label{eq:nc_strength}
}
We could explain this interesting effect based on the extrem order statistics of the strength 
of single fibers, i.e.\ we pointed out that the bundle strength increases until the strongest 
fiber dominates the ultimate failure of the system \cite{kadar_pre_2017}.
For sufficiently small systems, at high cutoffs $\varepsilon_{max}$, the strongest fiber 
can be so strong that it can keep the entire load on the system. Beyond the characteristic
system size $N_c$, this is no longer possible  so that the average strength decreases with $N$.
In the following we show that the fat tailed disorder of fibers' strength gives rise also to a 
complex behaviour of the statistics of breaking bursts when the parameters $\mu$ and 
$\varepsilon_{max}$ are varied.

\section{Statistics of breaking bursts}
Inside the quasi-brittle phase, we analyse the fracture process of the bundle under 
quasi-static loading, which is realized by slowly increasing the external load 
to provoke the breaking of a single fiber at a time.
For simplicity, we assume that 
the load of the broken fiber is equally redistributed over the intact ones which 
may trigger additional breakings, followed again by load redistribution. 
\begin{figure}
\begin{center}
\epsfig{bbllx=100,bblly=30,bburx=770,bbury=650, file=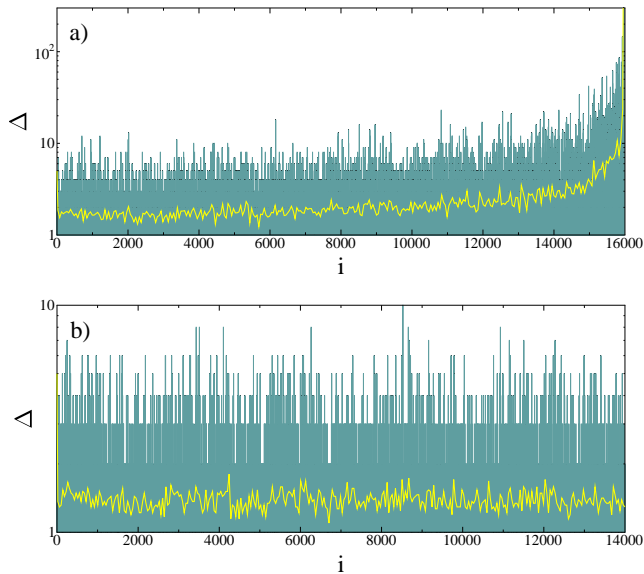, width=8.5cm}
  \caption{Series of bursts in a small system of $N=10^5$ fibers at the exponent $\mu=0.8$
  for two different values of the upper cutoff $(a)$ $\lambda=100$, and $(b)$ 
  $\lambda=+\infty$. The size of bursts $\Delta$ is presented as a function of the 
  order number $i$ of events.
  The yellow lines represent the moving average of burst sizes $\Delta$
  averaging over 25 consecutive data points.
  \label{fig:timeseries}}
\end{center}
\end{figure}
As a consequence of the repeated breaking and load redistribution 
steps, an avalanche emerges which stops when all the remaining intact
fibers are sufficiently strong to keep the elevated load. 
Global failure occurs 
in the form of a catastrophic avalanche which destroys the entire system.
The size $\Delta$ of the avalanche is defined as the number 
of fibers breaking in the correlated trail. 

\subsection{Acceleration towards failure}
\label{sec:a}
Inside the brittle phase (see Fig.\ 
\ref{fig:crack}) already the first avalanche triggers the immediate catastrophic failure of the 
system. However, in the quasi-brittle parameter regime the system gradually approaches 
failure through a sequence of bursts whose size $\Delta$ spans a broad range.
\begin{figure}
\begin{center}
\epsfig{bbllx=30,bblly=20,bburx=370,bbury=330, file=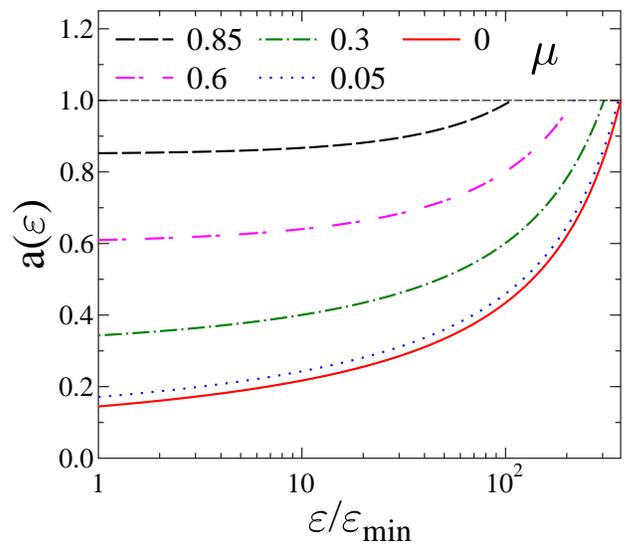, width=8.5cm}
  \caption{The average number of breaking fibers $a(\varepsilon)$ Eq.\ (\ref{eq:a}) 
  triggered by the failure of one 
  fiber due to the increase of the external load for several values 
  of the disorder exponent $\mu$. The cutoff strength $\varepsilon_{max}$ is fixed 
  to $\varepsilon_{max}/\varepsilon_{min}=1000$. All curves are presented 
  from $\varepsilon_{min}$ to the corresponding value of 
  $\varepsilon_c(\mu,\varepsilon_{max})$.
  For $\mu\to 0$ the critical point converges to $\varepsilon_c=\varepsilon_{max}/e$. 
  \label{fig:trogprob}}
\end{center}
\end{figure}
Representative examples of the series of bursts are shown in 
Fig.\ \ref{fig:timeseries} for two different values of the upper 
cutoff $\lambda=100$, $\lambda=+\infty$ at the same exponent $\mu=0.8$. 
For the infinite cutoff in 
Fig.\ \ref{fig:timeseries}$(b)$ the burst size $\Delta$ fluctuates, however, 
its moving average remains practically constant. It means that in spite of the 
increasing external load the system does not show any acceleration towards failure.
In fact, in this case the constitutive curve of the bundle (see Fig.\ \ref{epsilon_sigma_k})
does not have a maximum, it monotonically increases until the last fiber breaks the 
bundle. Contrary, for a finite upper cutoff in Fig.\ \ref{fig:timeseries}$(a)$
the system approaches global failure through an increasing average size of bursts.
At the critical point of failure a catastrophic avalanche emerges, while the catastrophic
event is absent when the cutoff strength is infinite.

To understand the behaviour of the burst sequence, it is instructive to calculate the average 
number $a$ of fiber breakings triggered immediately by the failure of a 
single fiber at the strain $\varepsilon$ \cite{hidalgo_avalanche_2009,kloster_burst_1997}. 
The load $\sigma = E\varepsilon$ dropped 
by the broken fiber is equally shared by the intact ones of number $N[1-P(\sigma)]$, 
giving rise to the stress increment $\Delta \sigma =\sigma/N[1-P(\sigma)]$. 
Multiplying $\Delta \sigma$ with the probability density
$p(E\varepsilon)$ of failure thresholds and with the total number of fibers $N$,
the average number of triggered breakings $a$ can be cast into the form
\begin{eqnarray}
a(\varepsilon) = \frac{E\varepsilon p(E\varepsilon)}{1-P(E\varepsilon)} = 
\frac{\mu}{1-\displaystyle{\left(\frac{\varepsilon}{\varepsilon_{max}}\right)^{\mu}}}.
\label{eq:a}
\end{eqnarray}
The right hand side of the equation was obtained by substituting the PDF $p$ Eq.\ 
(\ref{surusegfv}) and the CDF $P$ Eq.\ (\ref{eq:distrib}) of failure thresholds of our model.
The expression has to be evaluated over the range 
$\varepsilon_{min}\leq \varepsilon\leq\varepsilon_c$ which is illustrated 
by Fig.\ \ref{fig:trogprob} for several values of the exponent $\mu$ at a fixed upper 
cutoff $\varepsilon_{max}=1000$.
It can be seen that as the system approaches the critical point of global
failure $\varepsilon_c$ Eq.\ (\ref{eq:crit_strain}), the value of $a$ 
increases to 1 indicating the acceleration of the failure process 
and the onset of the catastrophic avalanche at the critical point. 

It follows from Eq.\ (\ref{eq:a}) that for an infinite upper 
cutoff $\varepsilon_{max}\to\infty$, the average number of triggered breakings $a$ 
takes a constant value $a=\mu<1$, which implies stable cracking 
and a constant average burst size as it could be inferred from 
Fig.\ \ref{fig:timeseries}$(b)$.
When the cutoff strength $\varepsilon_{max}$ is finite, 
for sufficiently small strains $\varepsilon$ the value of $a$ still can be considered 
constant $a\approx \mu$ and the acceleration of the bursting process is constrained to
the vicinity of the critical point $\varepsilon_c$. Eq.\ (\ref{eq:a}) implies that 
the effect is more pronounced when $\varepsilon_c\ll\varepsilon_{max}$ which requires
$\mu$ to be close to 1 and a large value of the cutoff strength according to 
Eq.\ (\ref{eq:crit_strain}).
Figure \ref{fig:trogprob} shows this behaviour for $\mu=0.85$ where $a$ 
remains close to $\mu$ for a broad range of $\varepsilon$, while for smaller 
exponents $\mu$ a considerable acceleration is observed from the beginning 
of the failure process. In the limit $\mu\to 0$ the number of triggered 
breakings takes the form
\beq{
a(\varepsilon)\approx \frac{1}{\ln{(\varepsilon_{max}/\varepsilon)}},
\label{eq:a_munull}
}
while the critical point $\varepsilon_c$ converges to $\varepsilon_c=\varepsilon_{max}/e$
(see also Fig.\ \ref{fig:trogprob}). 

Note, however, that in the derivation of $a$ implicitly an infinite system size 
is assumed. Later on we show that to obtain acceleration towards 
failure and a catastrophic avalanche at finite 
cutoff strengths, the size of the system $N$ has to exceed a characteristic 
value, which is a consequence of the fat tailed disorder.
\begin{figure}
\begin{center}
\epsfig{bbllx=30,bblly=30,bburx=380,bbury=330, file=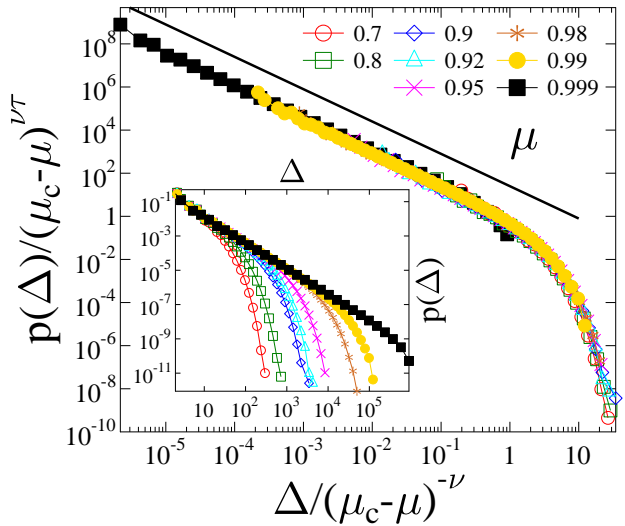, width=8.3cm}
  \caption{Inset: Size distribution of bursts $p(\Delta)$ for a bundle of $N=10^6$ fibers at 
  several values of the disorder 
  exponent $\mu$ when the cutoff strength of fibers is infinite $\varepsilon_{max}=+\infty$.
  Main panel: data collapse of the curves of the inset obtained by rescaling with a power 
  of the distance from the critical point $\mu_c=1$. Along the horizontal axis the scaling exponent
  is $\nu=2$ in agreement with Eq.\ (\ref{eq:powdiv}), while along the vertical axis
  the product $\nu\tau$ is used with $\tau=3/2$. 
  The straight line represents a power law of exponent $3/2$. \label{fig:inf_cut}}
\end{center}
\end{figure}

\subsection{Size distribution of bursts}
The statistics of breaking bursts can be characterized by the distribution 
$p(\Delta)$ of their size $\Delta$. 
The complete size distribution $p(\Delta)$ can be obtained analytically 
by substituting $a(\varepsilon)$ into the generic form 
\cite{hemmer_distribution_1992,kloster_burst_1997,hidalgo_avalanche_2009}
\begin{eqnarray}
&& \frac{p(\Delta)}{N} = \label{eq:sizedist}
\\ && \frac{\Delta^{\Delta - 1}e^{-\Delta}}{\Delta !}\int_0^{x_c}
p(x)a(x)\left[1-a(x)\right]^{\Delta - 1}e^{\Delta a(x)}dx, \nonumber
\end{eqnarray}
where for the upper limit of integration $x_c$ we have to insert the strength of
the bundle. Utilizing the approximation 
$\Delta ! \simeq \Delta^{\Delta}e^{-\Delta}\sqrt{2\pi\Delta}$, in
\begin{figure}
\begin{center}
\epsfig{bbllx=30,bblly=40,bburx=380,bbury=330, file=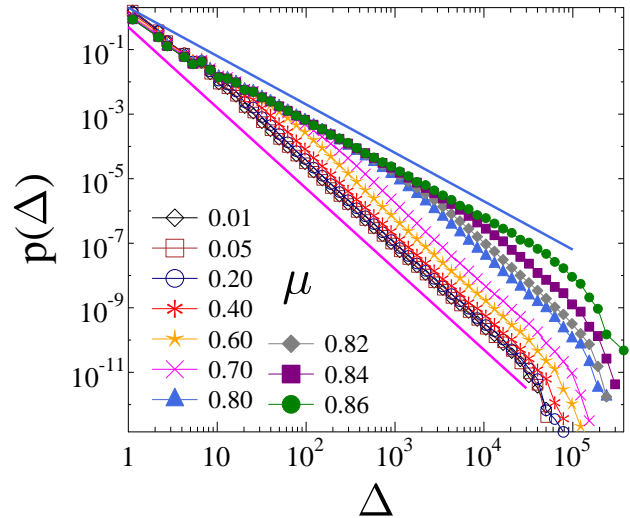, width=8.3cm}
  \caption{Burst size distributions $p(\Delta)$ in a bundle of size $N=10^7$
  at a fixed upper cutoff $\varepsilon_{max}=10$
  varying the value of the exponent $\mu$. The two straight lines represent power laws of 
  exponent $3/2$ and $5/2$.
   \label{fig:burstsize_mu}}
\end{center}
\end{figure}
the limiting case of an infinite upper cutoff with $a(\varepsilon)=\mu$ 
the burst size distribution can be cast into a simple analytic form
\begin{eqnarray}
\frac{p(\Delta)}{N} \simeq \Delta^{-\tau}e^{-\Delta/\Delta^*}.
\label{eq:burst_dist_mu}
\end{eqnarray}
A power law of exponent $\tau=3/2$ is obtained followed by an exponential cutoff. 
Here, $\Delta^*$ denotes the characteristic
burst size, which controls the cutoff of the distribution
\begin{eqnarray}
\Delta^* = \frac{1}{\mu -1 -\ln\mu}.
\label{eq:delta_c}
\end{eqnarray}
This result means that at an infinite upper cutoff of fiber strength 
$\varepsilon_{max}=+\infty$ the size distribution of bursts always follows 
a simple power law of a universal 
exponent $\tau=3/2$, where the value of the disorder exponent $\mu$ only controls the 
cutoff burst size $\Delta^*$. Using the Taylor expansion of logarithm around 1,
it can easily be shown that as $\mu\to \mu_c=1$ the cutoff burst size has 
a power law divergence
\beq{
\Delta^* \sim (\mu_c-\mu)^{-\nu},
\label{eq:powdiv}
}
with a universal exponent $\nu=2$. Burst size distributions obtained by computer simulations
of a bundle of size $N=10^6$ fibres are presented in the inset of 
Fig.\ \ref{fig:inf_cut} for
several $\mu$ values using an infinite cutoff strength. 
An excellent 
agreement is obtained with the analytical predictions. The main panel of 
Fig.\ \ref{fig:inf_cut} demonstrates that rescaling the distributions with 
$(\mu_c-\mu)^{-\nu}$ the curves of different $\mu$ can be collapsed 
on the top of each other which confirms 
the validity of the scaling law Eq.\ (\ref{eq:powdiv}). In Ref.\ 
\cite{danku_disorder_2016} we also showed that approaching $\mu_c=1$ at $\varepsilon_{max}=+\infty$, 
a continuous phase transition emerges from the quasi-brittle to the brittle phase, 
and we determined the critical exponents of the transition.
Note that the modified gamma form of Eq.\ (\ref{eq:burst_dist_mu}) of the burst size distribution
has also been proposed for earthquake magnitude distributions 
to maintain a finite strain release rate in natural earthquake populations
\cite{main_burton_bull_1984,kagan_geophys_1991}.

To characterize the statistics of breaking bursts at finite cutoff strength $\varepsilon_{max}$,
we determined the burst size distribution $p(\Delta)$ for several parameter sets 
along two straight lines inside the quasi-brittle phase of the bundle (see 
Fig.\ \ref{fig:crack}).
Figure \ref{fig:burstsize_mu} presents $p(\Delta)$ varying the 
disorder exponent $\mu$ at a constant finite upper cutoff $\varepsilon_{max}$.
It can be seen that approaching the phase boundary $\mu\to \mu_c(\varepsilon_{max})$ 
the burst size distribution tends to a power law functional form 
followed by an exponential cutoff consistent with the generic expression 
Eq.\ (\ref{eq:burst_dist_mu}).
\begin{figure}
\begin{center}
\epsfig{bbllx=30,bblly=40,bburx=380,bbury=330, file=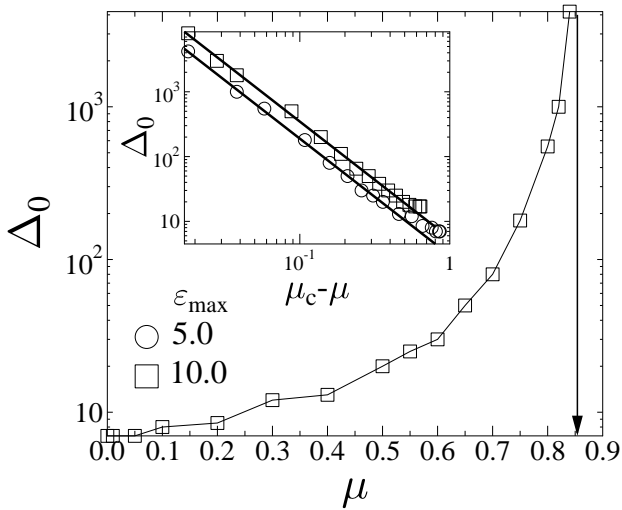, width=8.3cm}
  \caption{Crossover burst size $\Delta_0$ as a function of the 
  disorder exponent $\mu$ for the cutoff strength $\varepsilon_{max}=10$. 
  The arrow indicates the position of the corresponding critical point $\mu_c$.
  The inset presents $\Delta_0$ as a function of the distance from the critical 
  point $\mu_c(\varepsilon_{max})-\mu$ for two upper cutoffs on a double 
  logarithmic plot.
   \label{fig:crossover}}
\end{center}
\end{figure}
The value of the power law exponent is the same $\tau=3/2$ as for an infinite cutoff.
As $\mu$ decreases from its critical value, the burst size distribution 
exhibits a crossover between two power law regimes, i.e.\ the power law of exponent 
$\tau=3/2$ is followed by a steeper one of exponent $\tau=5/2$ in the regime 
of large bursts. For decreasing $\mu$ the crossover burst size $\Delta_0$ 
separating the two power law regimes, shifts to lower values. In the limit $\mu\to 0$ 
almost the complete size distribution can be described by a single power law 
of exponent $5/2$, however, the crossover burst size takes a small but finite 
minimum value. 

For moderate amount of disorder, it has been shown for 
fiber bundles under equal load sharing conditions that the 
size distribution of avalanches has a power law functional form
with a universal exponent $\tau=5/2$ \cite{kloster_burst_1997}. 
The result proved to be valid 
for those threshold distributions extending down to zero strength and having 
a sufficiently fast decreasing tail, where the constitutive curve 
$\sigma(\varepsilon)$ has a quadratic maximum 
\cite{hemmer_distribution_1992,kloster_burst_1997}.
In our system, the reason for the crossover of the burst size distribution $p(\Delta)$
between two power 
laws of exponent $3/2$ and $5/2$ is that the lower bound of fibers' 
strength $\varepsilon_{min}$ has a finite non-zero value. Additionally,
close to the boundary of the quasi-brittle phase, 
bursts are generated in a narrow strain interval since the breakdown point 
$\varepsilon_c$ falls close to $\varepsilon_{min}$.
\begin{figure}
\begin{center}
\epsfig{bbllx=30,bblly=40,bburx=380,bbury=330, file=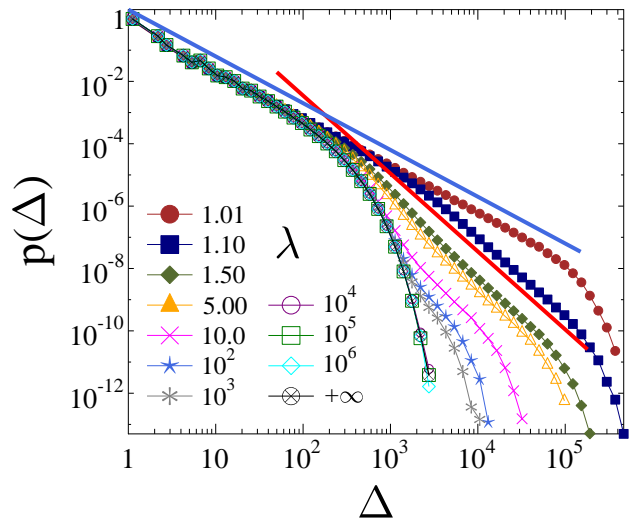, width=8.3cm}
  \caption{Burst size distributions $p(\Delta)$ for a fixed value of the exponent $\mu=0.85$
  varying the upper cutoff $\lambda$. The bundle is composed of $10^7$ fibers.
  The two straight lines represent power laws of exponent 
  $3/2$ and $5/2$. The case of an infinite upper cutoff $\lambda=+\infty$ 
  is also included for comparison.
  \label{fig:burstsize_lambda}}
\end{center}
\end{figure}
It was pointed out in Refs.\ 
\cite{pradhan_crossover_2005-1,pradhan_crossover_2006} that in such cases 
the crossover burst size $\Delta_0$ can be obtained as
\beq{
\Delta_0 = \frac{2}{a^{\prime}(\varepsilon_c)(\varepsilon_c-\varepsilon_{min})^2},
}
where $a^{\prime}(\varepsilon_c)$ denotes the derivative of $a(\varepsilon)$ at the 
breakdown point. To apply this generic result to our truncated fat tailed disorder
distribution, we substitute Eqs.\ (\ref{eq:crit_strain},\ref{eq:a})
which yields
\beq{
\Delta_0 = \frac{2\varepsilon_{max}(1-\mu)^{1/\mu-1}}{
\left[\varepsilon_{max}(1-\mu)^{1/\mu}-\varepsilon_{min}\right]^2}.
\label{eq:delta_cross}
}
This expression is valid for exponents $0<\mu\leq \mu_c(\varepsilon_{max})$. 
It can be seen in Eq.\ (\ref{eq:delta_cross}) that approaching the phase boundary
$\mu\to\mu_c(\varepsilon_{max})$, the crossover size diverges $\Delta_0\to +\infty$,
and hence, the burst size distribution $p(\Delta)$ 
has a single power law regime of exponent $\tau=3/2$.
The crossover to a higher exponent $\tau=5/2$ for large bursts
is observed away from the phase boundary where 
$\Delta_0$ takes finite values (see Fig.\ \ref{fig:burstsize_mu}).
Starting from Eq.\ (\ref{eq:delta_cross}), it can simply be shown that the 
divergence is described by a power law
\beq{
\Delta_0 \sim (\mu_c-\mu)^{-\gamma},
\label{eq:delta_c_div}
}
with a universal exponent $\gamma=2$. To test the validity of this prediction 
Eq.\ (\ref{eq:delta_c_div}),
we determined the value of $\Delta_0$ numerically as the crossing point 
of fitted power laws of exponents $3/2$ and $5/2$.
Figure \ref{fig:crossover} demonstrates that the
crossover burst size $\Delta_0$
rapidly increases as $\mu_c$ is approached, and it has a power law dependence on the 
distance from the critical point $\mu_c-\mu$, in agreement with Eq.\ (\ref{eq:delta_c_div}).
The exponent of the fitted power law is $\gamma=1.87\pm 0.1$ which falls close
to the analytical prediction.
\begin{figure}
\begin{center}
\epsfig{bbllx=30,bblly=40,bburx=380,bbury=330, file=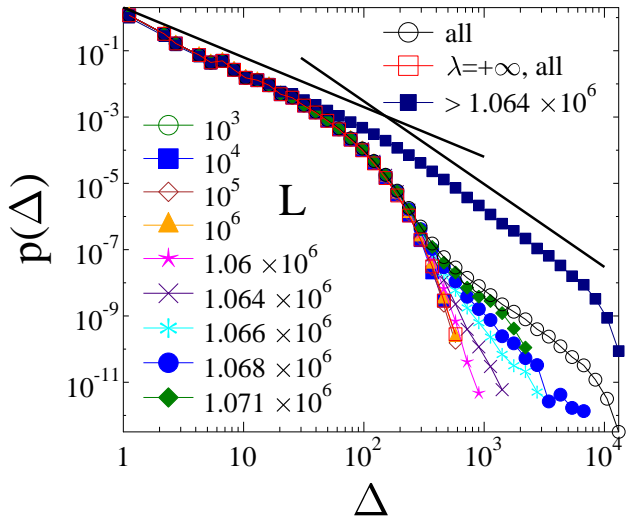, 
width=8.3cm}
  \caption{Size distributions $p(\Delta)$ of the first
  $L$ bursts of a bundle of $N=10^7$ fibers varying $L$ in a broad range at the 
  fixed disorder parameters $\mu=0.8$ and $\lambda=500$.
  The complete distribution of the entire failure process is also presented
  together with the distribution of the last events just preceding global failure.
  The total number of events is about $1.08\times 10^6$. The two straight lines
  represent power laws of exponents $3/2$ and $5/2$. 
  \label{fig:burstsize_lambda:box}}
\end{center}
\end{figure}

\section{Size dependent avalanche statistics}
When the cutoff strength $\varepsilon_{max}$ 
is varied while keeping the disorder exponent $\mu$ fixed,
the burst size distribution exhibits an even more complicated behaviour.  
For a fixed $\mu$, we express the cutoff strength relative to the phase boundary 
using the parameter $\lambda=\varepsilon_{max}/\varepsilon_{max}^c$, which takes
values in the range $\lambda>1$.
Figure \ref{fig:burstsize_lambda} presents $p(\Delta)$ for several values of 
$\lambda$ at the disorder exponent $\mu=0.85$, i.e.\ along the 
vertical dashed line inside the quasi-brittle phase of Fig.\ \ref{fig:crack}. 
It can be observed that starting from a single power law 
of exponent $\tau=3/2$ at the phase boundary,   
$p(\Delta)$ shows again a crossover between two power 
law regimes, where the crossover burst size $\Delta_0$ 
shifts to lower values as $\lambda$ increases. Starting from Eq.\ (\ref{eq:delta_cross})
it is easy to show that  $\Delta_0$ exhibits again a power law divergence 
\beq{
\Delta_0\sim (\lambda-1)^{-\gamma},
}
when approaching the phase boundary $\lambda\to 1$. The value of the exponent $\gamma$
is the same $\gamma=2$ as in Eq.\ (\ref{eq:delta_c_div}).
However, a significant difference, compared to the case of a constant cutoff, 
is that far from the phase boundary, after some transients, 
the steeper power law regime of exponent $\tau=5/2$ gradually disappears.
A single power law remains with exponent $\tau=3/2$, as at the phase boundary
Eq.\ (\ref{eq:burst_dist_mu}), but with a significantly lower cutoff burst 
size $\Delta^*$. 
\begin{figure}
\begin{center}
\epsfig{bbllx=40,bblly=20,bburx=370,bbury=330, file=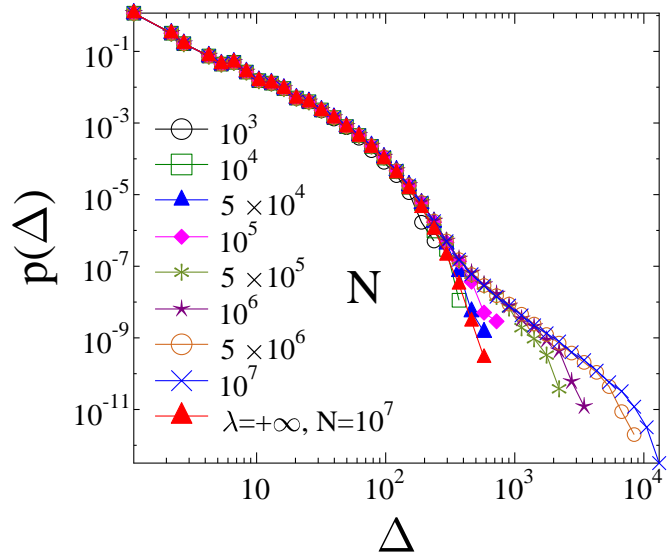, 
width=8.3cm}
  \caption{Size distribution of bursts $p(\Delta)$ for different system sizes $N$ 
  at fixed values of the disorder parameters $\mu=0.8$ and $\lambda=500$. 
  For small system sizes $p(\Delta)$ agrees 
  with the corresponding distribution of the infinite cutoff $\lambda=+\infty$.
  Above a characteristic system size a second power law regime gradually develops 
  for large bursts.
  \label{fig:burst_systemsize}}
\end{center}
\end{figure}

It is important to note in Fig.\ \ref{fig:burstsize_lambda} 
that at sufficiently large cutoffs $\lambda>1000$, the burst size 
distributions coincide with the one corresponding to the infinite cutoff $\lambda=+\infty$,
in spite of the fact that the system has a finite critical point $\varepsilon_c$.
The reason is that, at the $\mu$ exponent considered, 
the beginning of the series of bursts is close to stationary as it has been 
illustrated in Fig.\ \ref{fig:timeseries}$(a)$. Since the average number of 
triggered breakings $a(\varepsilon)$ is nearly constant over a broad 
range of strain $\varepsilon$, as $\lambda$ increases, the critical point 
is preceded by a shorter and shorter accelerating regime which has 
a diminishing contribution to the entire distribution $p(\Delta)$. 

To test this idea we analyzed in detail the statistics of burst sizes in a bundle
of size $N=10^7$ at the disorder parameters $\mu=0.8$ and $\lambda=500$ 
where both power law regimes are present. 
Figure \ref{fig:burstsize_lambda:box} shows 
the burst size distribution $p(\Delta, L)$  evaluated 
in event windows containing the first $L$
bursts, i.e.\   $p(\Delta, L)$ is the size distribution
of bursts $\Delta_i$, $i=1,\ldots , L$, averaged over several realizations of the disorder
at a given value of $L$.
For comparison, the size distribution of the entire failure process is also presented
together with the one corresponding to the case of 
an infinite cutoff $\lambda=+\infty$ obtained at the same system size $N$ and $\mu$ exponent.
It can be seen that up to the first $L\approx 10^6$ bursts, the distributions $p(\Delta, L)$ 
perfectly agree with the case of an infinite cutoff $p(\Delta,\lambda=+\infty)$. 
Deviations from $p(\Delta,\lambda=+\infty)$ start
around $L\approx 1.06\times 10^6$ above which gradually a steeper power law regime develops.
The result confirms that in spite of the existence of a well defined
critical point $\varepsilon_c$, for a broad event range the statistics of burst sizes
is consistent with the stationary process of the infinite strength cutoff, 
and acceleration towards failure is restricted to the close vicinity
of $\varepsilon_c$. The argument is further supported by the size distribution
of the last bursts with event index greater than $L=1.064\times 10^6$, 
which are generated in the vicinity of global failure. 
In this regime the functional form of $p(\Delta)$ is consistent with what has been 
obtained for varying $\mu$ in Fig.\ \ref{fig:burstsize_mu}, i.e.\ a crossover emerges
between two power laws of exponents $\tau=3/2$ and $\tau=5/2$, as expected in 
the vicinity of the critical point.
\begin{figure}
\begin{center}
\epsfig{bbllx=40,bblly=20,bburx=370,bbury=330, file=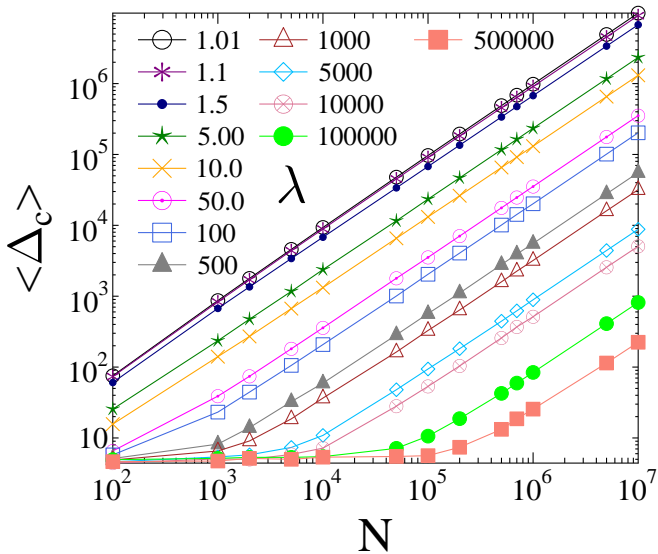, 
width=8.3cm}
  \caption{Average size of the catastrophic avalanche $\left<\Delta_c\right>$
  as a function of the system size $N$ for several values of the upper cutoff $\lambda$ 
  of fibers strength at a fixed exponent $\mu=0.8$.
  \label{fig:catast_burst_systemsize}}
\end{center}
\end{figure}

In Ref.\ \cite{kadar_pre_2017} we have shown that for fat tailed distributions of 
fiber strength, the number of fibers $N$ has a substantial effect on the ultimate 
failure strength of the bundle: for small system sizes the strongest fiber controls 
the macroscopic failure, consequently the average bundle strength increases with the 
system size $N$ described by Eq.\ (\ref{eq:strength_size}). 
The number of fibers $N$ has to exceed a characteristic value 
to observe the usual decreasing trend towards the strength 
of the infinite system given by Eqs.\ (\ref{eq:crit_strain},\ref{eq:sigma_c}).
Since at large $\lambda$ the system size $N$ controls the behaviour 
of the system at the critical point, it
follows that $N$ must play a decisive role also for the statistics 
of breaking avalanches. This is illustrated in Fig.\ \ref{fig:burst_systemsize} which
presents burst size distributions of bundles of different sizes $N$ at fixed values 
of the disorder parameters $\mu=0.8$ and $\lambda=500$. It can be observed that 
for small $N$ values, the burst size distributions $p(\Delta)$ coincide with the 
corresponding curve of a large system $N=10^7$ obtained at 
the infinite cutoff $\lambda=+\infty$. Above the
system size $N\approx 10^5$ a second power law regime gradually develops as it has been observed 
in Fig.\ \ref{fig:burstsize_lambda:box} for a single system size $N=10^7$ with varying 
event window $L$. 
\begin{figure}
\begin{center}
\epsfig{bbllx=40,bblly=20,bburx=370,bbury=330, file=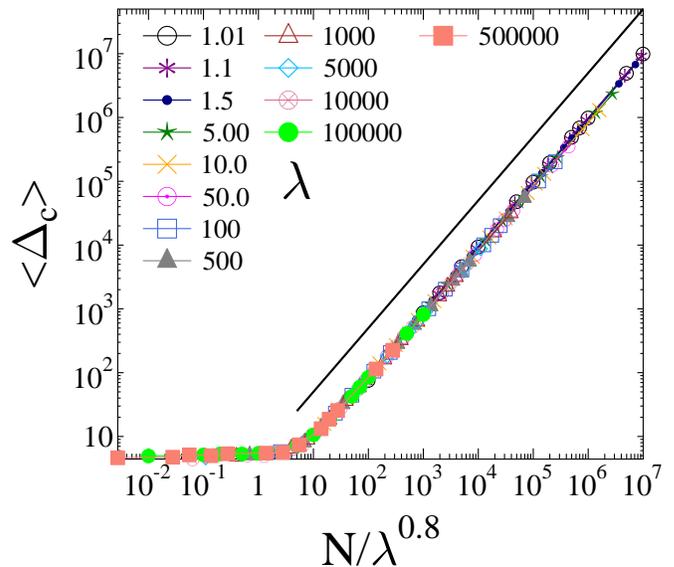, 
width=8.3cm}
  \caption{The same data as in Fig.\ \ref{fig:catast_burst_systemsize} is presented
  in such a way that along the horizontal axis the system size $N$ is rescaled with
  $\lambda^{\mu}$. High quality data collapse is obtained. The straight line represents 
  a power law of exponent 1.
  \label{fig:catast_burst_systemsize_scaled}}
\end{center}
\end{figure}

The reason of this astonishing dependence of the statistics
of avalanches on the size of the system is that for small system sizes,
even for finite cutoff strength of fibers, global failure occurs when the strongest 
fiber breaks. Consequently, the entire sequence of bursts is close to stationary and their 
statistics is practically the same as for the infinite cutoff. 
The existence of a finite critical point $\varepsilon_c$ is only realized when the system 
size $N$ exceeds a characteristic value $N_c$. For bundles with $N>N_c$ global failure 
is preceded by an acceleration of the failure process with increasing burst sizes.
In this regime macroscopic failure occurs in the form of a catastrophic avalanche, 
however, the catastrophic event is completely absent for $N<N_c$.
In order to quantify this crossover of the avalanche statistics with respect to the 
size of the system $N$, we determined the average size of the catastrophic avalanche 
$\left<\Delta_c\right>$ as a function of the size of the bundle $N$ varying the upper 
cutoff of fibers' strength $\lambda$ in a broad range.
The size of the catastrophic avalanche can be estimated as 
\beq{
\left<\Delta_c\right> \sim N\left(1-P(\varepsilon_c)\right),
}
so that if a well defined critical bundle strength $\varepsilon_c$ exists,
a linear dependence is obtained on the system size $\left<\Delta_c\right>\sim N$. 
Figure \ref{fig:catast_burst_systemsize} 
shows that for low $\lambda$ values the simulation results are consistent with the above 
prediction. However, far from the phase boundary $\lambda>1000$, 
a more complex behaviour is obtained: for small 
system sizes $\left<\Delta_c\right>$ does not depend on $N$, it takes a small constant 
value $\left<\Delta_c\right>\approx 7$. The regular linear increase with $N$ is restored 
above a characteristic system size $N_c$ which increases with $\lambda$.
Figure \ref{fig:catast_burst_systemsize_scaled} demonstrates that rescaling $N$
with the $\mu$th power of $\lambda$, the curves of $\left<\Delta_c\right>$ obtained at 
different $\lambda$ values can be collapsed on the top of each other. The high quality 
data collapse implies that the characteristic system size $N_c$, separating the two types of 
avalanche statistics, has a power law dependence on $\lambda$ as
\beq{
N_c \sim \lambda^{\mu}.
\label{eq:aval_nc}
}
This characteristic value $N_c$ is of course the same as the one which controls the size
scaling of the ultimate strength of the bundle Eq.\ (\ref{eq:nc_strength}) 
\cite{kadar_pre_2017}.
It also follows that the event window analysis presented in Fig.\ 
\ref{fig:burstsize_lambda:box}, can only be performed for system sizes $N>N_c$, and 
the crossover event index $L_c$ below which the burst size distribution is close to the 
one of the infinite cutoff, has the same dependence Eq.\ (\ref{eq:aval_nc}) on the 
disorder parameter.

\section{Discussion}
The degree of materials disorder has a substantial effect on the fracture of heterogeneous
materials both on the micro- and macro-scales. When subject to a slowly increasing external 
load, fracture proceeds in bursts which can be considered as precursors of global failure.
Failure forecast methods of the imminent catastrophic failure strongly rely on the 
bursting dynamics \cite{voight_relation_1989,tarraga2008447,vasseur_scirep_2015}. It has been 
demonstrated experimentally that increasing amount of disorder gives rise to 
a more intensive precursory activity which then improves the quality of forecasts 
\cite{vasseur_scirep_2015,salje_main_minecollapse_2017}.

In this paper we investigated the effect of the amount of disorder 
on the microscopic dynamics of the fracture process of heterogeneous materials 
in the framework of a fiber bundle model focusing on the limit of very high disorder. 
We considered a power law distribution of fibers' 
strength where the degree of disorder could be controlled by tuning the power law 
exponent and the upper cutoff of breaking thresholds. Assuming equal load sharing 
after local breakings, we showed that on the
macro-scale the mechanical response of the bundle is either perfectly brittle
where the bundle abruptly fails right at the breaking of the first fibre, or it is 
quasi-brittle where macroscopic failure is approached through a sequence of breaking
bursts.The evolution of the crackling event series and the statistics of burst sizes
have a high importance for the forecasting of the imminent failure of the bundle.

We showed that for an infinite upper cutoff of fibers' strength, 
the sequence of bursts is stationary in the sense that the average burst size 
is constant. Hence, the system does not exhibit any sign of acceleration towards 
failure. Consequently, a power law burst size distribution is obtained, where the disorder 
exponent only controls the cutoff burst size. 
For finite upper cutoffs we showed that there exists a well-defined critical point of global
failure, however, it can only be realized in sufficiently large systems. In small 
systems the global strength of the bundle is controlled by the strongest fiber.
This peculiar behaviour gives rise to an astonishing dependence of the statistics
of burst sizes on the size of the system: for small systems the burst sequence proved to be
close to stationary, and hence, the burst size distribution coincides with the one 
corresponding to the infinite upper cutoff of fibers' strength. For  large 
systems the initially stationary sequence is followed by an accelerating regime 
in the close vicinity of the critical point, which gives rise to a crossover between 
two power laws of the burst size distribution. Analysing the dependence of the 
average size of the catastrophic burst on the size of the bundle, we pointed out that
the transition between the two types 
of burst size distributions occurs at a characteristic system size which depends 
on the disorder parameters of the bundle. The results can have relevance for the design of 
laboratory experiments: when the micro-scale materials disorder has a rapidly (exponentially)
decaying distribution, the sample size mainly affects the cutoff of the size distribution
of bursts but not its functional form. However, for fat tailed disorder the sample size 
has a strong effect on the functional form of the burst size distribution so that the 
size of specimens in laboratory tests has to be sufficiently large to reproduce the acceleration of the 
burst sequence towards failure obtained in field measurements.

We also demonstrated that for a moderate amount of disorder, i.e.\ 
varying the disorder parameters in the vicinity 
of the phase boundary between the brittle and quasi-brittle phases, 
a crossover occurs between two power laws of exponents $3/2$ and $5/2$.
The reason is that bursts are generated in a narrow strain interval close to
the critical point of macroscopic failure. In this case the crossover burst size
was found to have a power law divergence as the phase boundary is approached.

Our results set important limitations on the forecastability of the imminent failure
\cite{voight_ffm_nature_1988,main_limits_2013,vasseur_scirep_2015}
of the system when the microscale disorder is fat tailed. We have demonstrated 
that even if a considerable avalanche activity accompanies the failure process, 
the collapse may not be predictable either because it is controlled by the extreme 
order statistics of fibers' strength, 
or the accelerating regime preceding failure is too short. 
In failure forecast methods accelerating precursors have to be identified above a null 
hypothesis of stationary event rate, then one needs to wait for a sufficient amount of data 
to define a singularity with accuracy and precision at a finite time before the time of ultimate
failure \cite{main_limits_2013}. 
The effect of high disorder on the statistics of breaking bursts, revealed by our study, 
may be a real limitation for practical applications 
of forecasting methods based on acoustic or seismic precursors of failure 
\cite{geller_forecast_science_1997,main_limits_2013}.

In the present study we focused
mainly on the integrated statistics of burst sizes considering all events up to failure.
The quantitative characterization of the evolution of the event series towards failure
requires further careful analysis which is in progress.

\begin{acknowledgments}
The work is supported by the EFOP-3.6.1-16-2016-00022 project. 
The project is co-financed by the European Union and the European Social Fund.
This research was supported by the National Research, Development and
Innovation Fund of Hungary, financed under the K-16 funding scheme Project no.\ K 119967. 
The research was financed by the Higher Education Institutional
Excellence Program of the Ministry of Human Capacities in Hungary, 
within the framework of the Energetics thematic
program of the University of Debrecen.
\end{acknowledgments}

\bibliography{/home/feri/papers/statphys_fracture}

\end{document}